\documentclass[manuscript]{acmart}

\AtBeginDocument{%
  \providecommand\BibTeX{{%
    \normalfont B\kern-0.5em{\scshape i\kern-0.25em b}\kern-0.8em\TeX}}}

\acmBooktitle{Proceedings of the 34th ACM Conference on Hypertext and Social Media (HT ’23),
 September 04--08, 2023, Rome, Italy} 
\acmPrice{15.00}
\acmISBN{978-1-4503-XXXX-X/18/06}

\copyrightyear{2023}
\acmYear{2023}
\setcopyright{acmlicensed}\acmConference[HT '23]{33rd ACM Conference on Hypertext and Social Media}{September 4--8, 2023}{Rome, Italy}
\acmBooktitle{33rd ACM Conference on Hypertext and Social Media (HT '23), September 4--8, 2023, Rome, Italy}
\acmPrice{15.00}
\acmDOI{10.1145/3603163.3609064}
\acmISBN{979-8-4007-0232-7/23/09}

\begin{document}

\title[The Looming Threat of Fake and LLM-generated LinkedIn Profiles]{The Looming Threat of Fake and LLM-generated LinkedIn Profiles: Challenges and Opportunities for Detection and Prevention}

\author{Navid Ayoobi}
\email{nayoobi@CougarNet.UH.EDU}
\author{Sadat Shahriar}
\email{sshahria@CougarNet.UH.EDU}
\author{Arjun Mukherjee}
\email{arjun@cs.uh.edu}

\affiliation{%
  \institution{University of Houston}
  \city{Houston}
  \state{Texas}
  \country{USA}
  \postcode{43017-6221}
}

\renewcommand{\shortauthors}{Ayoobi, \textit{et al.}}

\begin{abstract}
In this paper, we present a novel method for detecting fake and Large Language Model (LLM)-generated profiles in the LinkedIn Online Social Network immediately upon registration and before establishing connections.
Early fake profile identification is crucial to maintaining the platform's integrity since it prevents imposters from acquiring the private and sensitive information of legitimate users and from gaining an opportunity to increase their credibility for future phishing and scamming activities.
This work uses textual information provided in LinkedIn profiles and introduces the Section and Subsection Tag Embedding (SSTE) method to enhance the discriminative characteristics of these data for distinguishing between legitimate profiles and those created by imposters manually or by using an LLM.
Additionally, the dearth of a large publicly available LinkedIn dataset motivated us to collect $3600$ LinkedIn profiles for our research.
We will release our dataset publicly for research purposes.
This is, to the best of our knowledge, the first large publicly available LinkedIn dataset for fake LinkedIn account detection.
Within our paradigm, we assess static and contextualized word embeddings, including GloVe, Flair, BERT, and RoBERTa.
We show that the suggested method can distinguish between legitimate and fake profiles with an accuracy of about $95\%$ across all word embeddings.
In addition, we show that SSTE has a promising accuracy for identifying LLM-generated profiles, despite the fact that no LLM-generated profiles were employed during the training phase, and can achieve an accuracy of approximately $90\%$ when only $20$ LLM-generated profiles are added to the training set.
It is a significant finding since the proliferation of several LLMs in the near future makes it extremely challenging to design a single system that can identify profiles created with various LLMs.
\end{abstract}




\maketitle

\section{Introduction}
The advent of Online Social Networks (OSNs) has dramatically revolutionized the way in which individuals communicate and exchange information.
LinkedIn as the most renowned OSN for professional networking brings a unique opportunity for individuals and companies to find jobs, develop their businesses, recruit, and pursue talent acquisition.
However, as LinkedIn's user base has grown, there has been a corresponding increase in the number of fake profiles that cause issues for genuine users, companies, and the OSN itself.
LinkedIn's detailed user profiles make it a perfect venue for imposters to reach their intended audience \cite{prieto2013detecting}.
In addition, LinkedIn's lack of verification has exacerbated the problem \cite{maxlinkedin}.
Consequently, fraudsters can create accounts with minimal expense to access a vast number of potential victims.

Fake profiles can be defined as accounts that misrepresent the profile owner or contain fraudulent information.
These accounts are created for a variety of reasons, such as to boost the number of employees of a company in order to appear more influential than it actually is, or to utilize a company's reputation in order to be considered a legitimate profile for further purposes like phishing, scamming, or disseminating misleading information to attract customers \cite{kondeti2021fake}.
The proliferation of fake accounts diminishes the platform's credibility by causing a negative user experience for genuine LinkedIn members.
If users believe that LinkedIn is inundated with fake accounts, they may be less inclined to join the network and less likely to trust the information they discover there.
It additionally harms the advertising and revenue streams of the OSN \cite{xiao2015detecting}.
Typically, advertisers use LinkedIn to target certain demographics and sectors, and fake profiles can skew this targeting.
This lowers engagement, click-through, and advertising return-on-investment.
Moreover, recruiters may waste their valuable time and resources sifting through fake profiles.
The talent pool on LinkedIn could also be misrepresented by fake profiles, leading to inaccurate assumptions about the job market.

The CAPTCHA and phone number requirements upon registration may deter some fake accounts, but they can be circumvented using automated tools and disposable phone numbers, or VOIP services, respectively.
On the other hand, OSN users typically avoid reporting fake accounts for a variety of reasons.
First, fraudulent accounts are hard to spot precisely.
Second, they have no incentive to report the accounts, and they prefer to merely cancel connection requests upon the identification of fraudulent accounts.
In addition, processing submitted reports is excessively time-consuming due to the large number of LinkedIn members.
As a result, imposters continue their malicious activities and can even forge more connections, making it more difficult to identify them as fake accounts.

In the near future, the use of Large Language Models (LLMs) to build fraudulent profiles will compound the issue for OSN platforms since it will be extremely challenging to identify these profiles.
LLMs have been trained on a large text corpus to produce texts that are often indistinguishable from human-written content.
By utilizing an LLM algorithm to produce content for several sections of a LinkedIn profile, e.g., About, Education, Experience, and Skill sections, it is considerably simpler for imposters to create profiles that seem authentic.
Furthermore, an LLM could be used to compose messages that the fake profile can send to other users in an attempt to establish connections.
These messages could be tailored to look more authentic by referencing the personal information of the target profile.
Therefore, there is a need to design an automatic method for detecting human-generated, as well as LLM-generated fake profiles to prohibit them from interacting with legitimate OSN users by devising proper precautions.

In order to design machine learning (ML)-based systems that can reliably detect these fake profiles, it is necessary to train the models with labeled data.
To the best of our knowledge, the only publicly available dataset for LinkedIn fake profile detection is the one suggested by \cite{adikari2020identifying}.
This dataset includes only $17$ fake profiles posing a major barrier for researchers in their efforts to devise an ML method for detecting fake profiles.  
The scarcity of data is due to the difficulties involved in manually spotting fake profiles since LinkedIn's strict policies and its implemented measures prevent web scraping on their website, which makes data collection a burdensome process \cite{adikari2020identifying}.
However, to conduct our research, we collected $2400$ LinkedIn profiles, of which $1800$ and $600$ are legitimate and fake LinkedIn profiles, respectively.
In addition, we utilized an LLM (ChatGPT) to build $1200$ profiles that can serve as the basis for detecting next-generation fake profiles in the future.
Our dataset has been collected over the course of nine months and only comprises information that can be seen by everyone prior to establishing connections.
All profiles in our dataset have been examined and validated by the authors in order to produce a reliable resource for future studies.
We release the collected dataset to the research community so that researchers can conduct further study on this topic. 

Utilizing numerical data to detect fake accounts in OSNs has been a common practice in prior research \cite{kaubiyal2019feature,adikari2020identifying,prieto2013detecting,cresci2015fame,ala2017spam,adewole2019smsad}. 
While these techniques are useful at identifying fake accounts, they frequently rely on network graph data or dynamic data, such as a user's activity, number of followers, and connections.
This fact imposes two challenges.
First, accessing dynamic data requires interaction with the fake accounts, allowing them access to the private and sensitive data of the legitimate accounts.
Second, collecting and manipulating dynamic data is a time-consuming process, giving fake accounts an opportunity to enhance their legitimacy before being detected as fake.
Therefore, the performance of current solutions degrades for detecting fake accounts immediately after registration and prior to establishing connections.
In addition, although several articles have investigated the detection of LLM-generated content \cite{salminen2022creating,mitchell2023detectgpt}, there are no approaches intended to identify fake accounts created by LLMs, to the best of our knowledge.

In this paper, we introduce the Section and Subsection Tag Embedding (SSTE) method for detecting LinkedIn fake accounts based on the textual data provided in the LinkedIn profiles.
We show that by subtracting the embeddings of section and subsection tags from the embedding representations of the provided textual data, we are able to increase the likelihood of differentiating fake profiles from legitimate profiles.
Our method is able to identify fake accounts immediately after user registration on the OSN and prior to establishing any connections with legitimate users.
We assess the efficacy of several word embeddings, including GloVe \cite{pennington2014GloVe}, Flair \cite{akbik2018contextual}, BERT-base \cite{devlin2018bert}, and RoBERTa \cite{liu2019roberta}, utilized in our SSTE technique for spotting fake LinkedIn accounts.
We show that the suggested method outperforms a model that uses solely numerical data by $17.79 \%$ in terms of accuracy.
In addition, we show that SSTE has an accuracy of about $70\%$ for identifying LLM-generated profiles, despite the fact that no LLM-generated profiles were employed during the training phase.
We also conduct an experiment using LLM-generated profiles instead of LinkedIn fake profiles as the fake samples in the training phase because finding and collecting LinkedIn fake accounts is extremely challenging.
We demonstrate that in this case our model performs reasonably well, and it shows a promising starting point for further refinement.
In addition, with the proposed technique, it is sufficient to include only a small number of LLM created profiles (about 20) in the training set in order to reach an accuracy of approximately $90\%$ in distinguishing legitimate profiles from fake and LLM-generated profiles.

The main contributions of this paper are summarized as follows.
\begin{itemize}
    \item We build and publish a reasonably large dataset for detecting fake LinkedIn profiles which consists of legitimate, fake and LLM-generated profiles.
    \item Our approach detects fake profiles  as quickly as feasible after registration without using dynamic data or connecting to the fake accounts.
    \item To the best of our knowledge, this is the first fake profile detector that is capable of discriminating legitimate profiles from fake profiles created by both humans and LLMs.
\end{itemize}\nointerlineskip
\section{Related work}
Despite extensive research on recognizing fake profiles in OSNs \cite{kaubiyal2019feature,agarwal2019analyzing,breuer2020friend,yuan2019detecting,wanda2020deepprofile}, there is a paucity of literature that concentrates on detecting LinkedIn fake profiles \cite{roy2020fake,adikari2020identifying,prieto2013detecting,xiao2015detecting,kontaxis2011detecting}. 
Generally the primary traits utilized by all of these fake account detectors can be grouped into static and dynamic (activity-based) data.
Static data are the information that do not change over time and are unaffected by a user's actions on the OSN. 
In contrast, dynamic data refer to the information that vary over time and are impacted by the user's actions on the OSN. This consists of the user's posting frequency, number of connections and interactions, and post content.

\subsection{Other OSNs}
Several studies have investigated the use of dynamic data for detecting fake accounts on Facebook and Twitter.
Kaubiyal and K. Jain in \cite{kaubiyal2019feature} proposed a method for detecting fake profiles in Twitter.
They utilized several dynamic data in their method such as count of retweets, count of hashtags and mentions, and the number of Tweets the user posts per day.
Then, they evaluated logistic regression (LR), support vector machine (SVM), and random forest (RF) classifiers on discriminating "bot" and "human" profiles using mentioned features.
Wani \textit{et al.} \cite{agarwal2019analyzing} presented a fake profile detection model using $12$ sentiment-based features extracted from Facebook accounts' posts.
The first eight features were based on Plutchik's eight fundamental emotions, while the ninth feature measured the variety of categories that individuals indicated in their postings.
The tenth feature represented the variance in the posts' emotions, and the remaining two features were related to the percentage of posts with positive and negative sentiments.
They achieved an accuracy of about $91\%$ using an RF classifier.
In \cite{breuer2020friend}, the authors introduced \textit{SybilEdge}, a graph-based method for detecting fake Facebook profiles in early stages.
An aggregation technique is adopted to assign higher weight to the selection of targets made by a user based on their popularity among fake users as opposed to real users, in addition to considering the response of these targets towards fake versus real users.
For spotting fake accounts the posterior probability is computed as a function of the user's set of friend request targets as well as the responses received from these targets.
As these methods rely on the historical data and user activities that are not yet available for newly registered accounts, they may yield inaccurate outcomes for identifying newly registered fake accounts. 

\subsection{LinkedIn platform}
Adikari and Dutta \cite{adikari2020identifying} performed a feature selection method using PCA on a set of features extracted from LinkedIn profiles to find the best set of features for discriminating real from fake profiles. 
They trained their model on a small dataset including $20$ and $17$ real and fake profiles, respectively.
They achieved an accuracy of $87.34\%$ by testing the trained model on the same volume of data as their training set.
Including number of connections and recommendations in selected features hinders the effectiveness of this method on identifying fake profiles immediately after registration. 
Prieto \textit{et al.} \cite{prieto2013detecting} proposed two detection methods for identifying spammers and spam nets.
They analyzed several features including the number of words in a profile, the number of contacts, the length of the profile name and location, the profile photo, and existence of plagiarism in the profile. 
They reported that spammer profiles are simpler and contains less details compared to legitimate profiles.
Unlike previous studies that focused on identifying fake profiles or bots, Xiao \textit{et al.} in \cite{xiao2015detecting} aimed to identify instances at registration time or shortly thereafter where a single user created a cluster of profiles on LinkedIn platform.
They firstly clustered raw list of accounts based on predefined parameters including cluster size, time span of registered accounts, and a criteria like similar IP addresses.
Then, a numerical vector representation is computed for each cluster using basic distribution features, pattern features, and frequency features extracted from the accounts within a cluster. 
These vectors are then fed to an LR, SVM, or an RF classifier to obtain the likelihood of being fake.
Kontaxis \textit{et al.} \cite{kontaxis2011detecting} proposed a method for detecting cloned profiles where attackers duplicate a user's profile in LinkedIn and other OSNs.
User-specific information is firstly extracted from the target legitimate profile.
Then, several queries base on this information are passed through a search engine. 
The pieces of information with fewer search engine results are selected as the user-identifying phrases.
The user's full name along with his/her identifying phrases are used to locate profiles that are potentially related to the user.
A similarity score is calculated based on the common values of information fields.
Additionally, they compared profile pictures of listed profiles with the profile picture of target account as cloned profiles tend to use the victim's picture to boost their credibility.

Our research introduces two significant innovations that distinguish it from previous works in LinkedIn fake account detection. 
Firstly, our method has the capability to identify fake profiles immediately after registration without the need for establishing connections or utilizing dynamic data. 
Secondly, our proposed method is able to identify fake accounts created by imposters both manually or by using an LLM.
\begin{figure}
    \centering
    \includegraphics[width=0.9\linewidth]{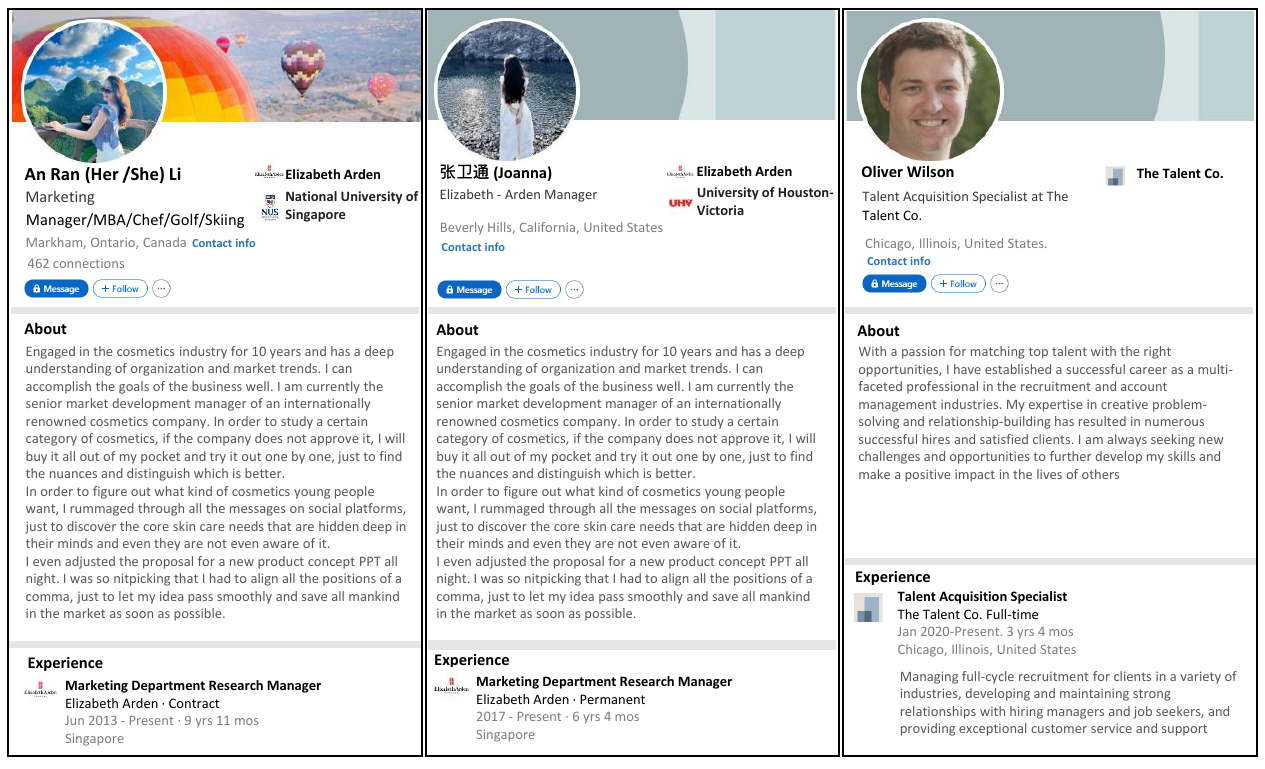}
    \caption{Examples of fake LinkedIn accounts. The left and middle profile are two examples of FLPs. Both used the same contents in the About and Experience sections and a non-professional photo. The right profile shows a fake profile created by an LLM (ChatGPT).}
    \label{fig:profiles}
\end{figure}
\section{Data collection}
We collected a dataset containing $3600$ profiles for our research.
The dataset consists of $1800$ legitimate LinkedIn profiles (LLPs), $600$ fake LinkedIn profiles (FLPs) and $1200$ profiles generated by ChatGPT (CLPs) to gain insight into potential future fake profiles created by LLMs.
The dataset only contains information that is accessible to every LinkedIn user prior to initiating a connection.
There are two justifications for this choice.
First, this study detects fake accounts created immediately after registration and before any connections are made to prevent imposters from gaining access to the information of real users.
Second, we cannot add information that is accessible only to a user's connected people due to privacy considerations.
The dataset includes the workplace, location, number of connections and followers, status of profile picture, and the information of various sections including About, Experiences, Educations, Licenses, Volunteers, Skills, Recommendations, Projects, Publications, Courses, Honors and Awards, Scores, Languages, Organizations, Interests, and Activities that are visible to all users.
In addition, columns with numerical attributes were added to our dataset representing the number of components in each section.
In this research, we omit some columns from the dataset as we intend to construct a detector for newly registered profiles.
However, we release the comprehensive dataset to the research community in order to enable scholars to explore new lines of inquiry and delve deeper into the topic.

In order to find FLPs, we searched hashtags like \#fake\_accounts, \#fake\_profiles, \#scammers ,\#spammers and \#bot, and were able to locate multiple LinkedIn posts complaining about fake accounts.
In addition, we collected some FLPs reported directly by LinkedIn users who received a large number of connection requests daily from fake accounts.
To ensure that the collected data is reliable and accurate, the authors manually reviewed these accounts.
Moreover, we discovered FLPs from organizations that attempted to enhance their personnel count by creating fake accounts.
We were able to locate these profiles by searching the job title and the name of the company using Bing search engine.
The left and middle profile in Fig. \ref{fig:profiles} show two examples of FLPs.
They used the same content in the About section, and the same job title and company name in the Experience section.
Additionally, they utilized a non-professional profile photo which is uncommon on the LinkedIn platform.

To reflect the future challenges in detecting fake profiles, we created $1200$ profiles with ChatGPT.
We hypothesized that individuals will use an LLM to complete the sections they were previously more likely to complete manually.
Thus, we began by sampling the profile statistics (the number of components for a particular section) from FLPs and LLPs.
Then, we supply ChatGPT with precise instructions to produce each section's information.
For example, we used following statement to generate three\footnote{The number is sampled from FLPs or LLPs statistics} components  for "Experiences" section:
\emph{"For the experience section of his/her Linkedin profile, generate 3 experiences containing his/her role, job title, name of the company, start and end date of this job and its duration, workplace location (including city, state, country), and a brief description about what he/she did in this position"}.
The right profile in Fig. \ref{fig:profiles} shows an example of fake profiles generated using ChatGPT. 
We used \textit{"facegen"} website \footnote{\url{https://facegen.io/}} to generate a fake profile photo for this profile for illustration purposes only.
The data collection procedure was completed over the course of nine months, starting in June 2022 and ending in February 2023.
The dataset will be released after publication \footnote{\url{https://github.com/navid-aub/LinkedIn-Dataset}}.

\section{Methodology}
\begin{figure}[t]
  \centering
  \includegraphics[width=0.95\linewidth]{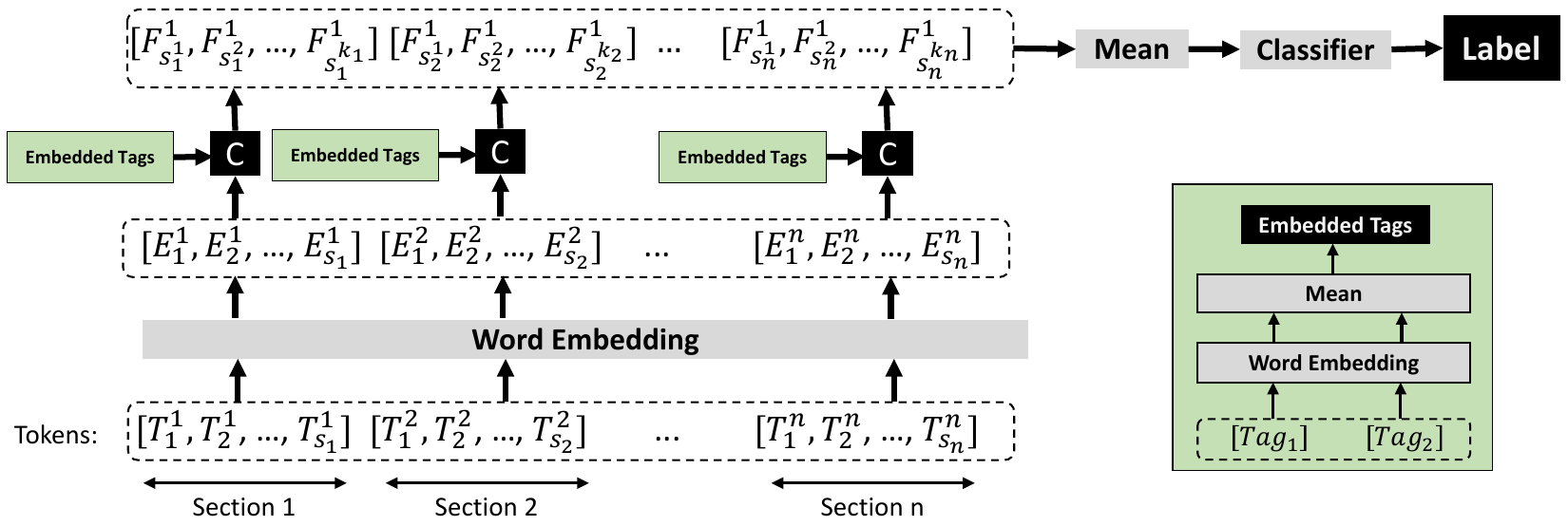}
  \caption{Overview of the proposed method}
  \Description{}
  \label{scheme}
\end{figure}

In this section, we provide an overview of the proposed method for detecting newly registered LinkedIn fake profiles.
In order to achieve this goal, we only use available information provided during the registration process to feed our classifier.
Therefore, we exclude the information like the number of followers, the number of connections, the information in recommendation, and activity sections as these require time to be formed.
An overview scheme of the proposed method is depicted in Fig. \ref{scheme}.

\subsection{Preprocessing}
To decrease the total vocabulary size and enhance the model's capacity to generalize to new data, all texts are converted to lowercase.
We then expand the contradictions to improve the consistency of the text among different sections.
The URLs, punctuation, stop words and white spaces are removed and accented characters are replaced with standard characters.
All numbers are written in words.
The texts are then split into a list of tokens and each token is reduced to its lemma using WordNetLemmatizer provided by Natural Language Toolkit (NLTK).

\subsection{Word embedding schemes}
In this research, we utilize four different types of word embeddings, namely GloVe \cite{pennington2014GloVe}, Flair \cite{akbik2018contextual}, BERT-base \cite{devlin2018bert}, and RoBERTa \cite{liu2019roberta} to represent the tokens as numerical vectors.

\subsubsection{GloVe embedding}
GloVe's primary objective is to generate a co-occurrence matrix that captures the frequency with which each word co-occurs with every other word in a large corpus of text. 
This matrix is then transformed into a word to word co-occurrence probability matrix where the matrix entries indicate the likelihood of detecting a word in the context of another word.
The GloVe then learns a word embedding for each word in the vocabulary, such that the dot product of the two word embeddings represents the co-occurrence probability of those two words.
GloVe not only relies on the words' local context information, but it also incorporates global statistics to capture both global and local semantic relationships between the words. 

\subsubsection{Flair embedding}
The fundamental goal of Flair is to produce contextual embeddings for words, which implies that a word's embedding is impacted by its surroundings in the text as well as by the word itself.
To achieve this goal, Flair employs a bidirectional language model processing text in both directions, i.e. from left to right and right to left.
In addition, Flair combines character-level and word-level representations to create its embeddings.
The word-level representations capture a word's semantics, whereas the character-level representations capture its morphological and syntactic characteristics.
Using character-level and word-level representations enables Flair to build embeddings that are resistant to terms not present in the lexicon, spelling changes, and uncommon phrases.

\subsubsection{BERT-base embedding}
The BERT generates contextualized word embeddings employing a $12$-layer transformer architecture with $110$ million parameters.
Each layer contains $12$ attention heads, and $768$ hidden units. 
The model is trained based on both the left-to-right and right-to-left context using a large corpus of textual data.
By feeding it pairs of sentences during training, it learns to estimate the likelihood of each word in the second sentence given the first sentence.
Through this procedure, the model is able to extract the contextual information of each word in a sentence, producing extremely powerful word embeddings.
\subsubsection{RoBERTa embedding}
RoBERTa is an enhanced variant of the BERT base model. 
RoBERTa leverages a larger corpus of data for pre-training, in addition to a dynamic masking technique preventing the model from retaining specific data patterns.
RoBERTa provides contextualized word embeddings similar to BERT, but with enhanced performance on a wide variety of NLP applications.

\subsection{Section and Subsection Tag Embeddings (SSTE)}
\begin{figure}[!th]
    \centering
    \includegraphics[width=0.9\linewidth]{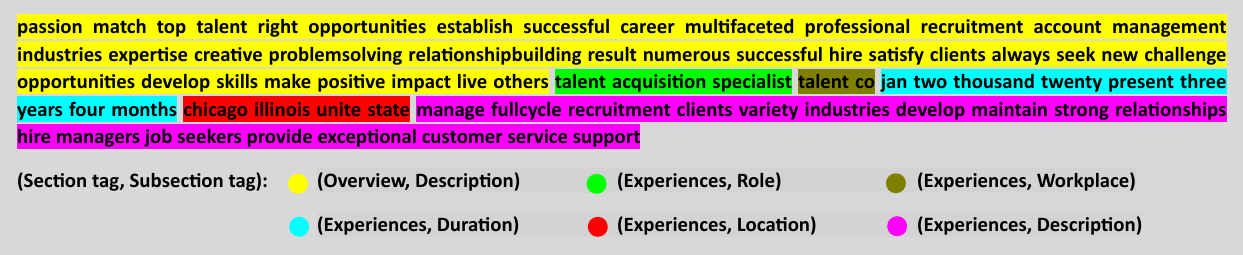}
    \caption{The concatenated and cleaned textual information of the CLP shown in Fig. \ref{fig:profiles}. The section and subsection tags for each word are specified by the same color as the word highlighted. The mean of section and subsection tag embeddings is subtracted from the word embedding to compute each word's final embedding. Then, the mean of final embeddings is computed to represent the whole document embedding.}
    \label{fig:con}
\end{figure}

The textual information provided in the various sections of a LinkedIn profile is concatenated to create a single document.
The produced document is then passed through the preprocessing module to arrive at a cleaned document.
The cleaned text is tokenized, and then fed to the word embedding function $Em(.)$.
\begin{equation}
    E_i^{j} = Em(T_i^{j}) \hspace{3em} i=1, ..., s_j
\end{equation}
where $T_i^j$ is the $i^{th}$ token, and $E_i^j$ is its embedded representation for the $j^{th}$ profile section. 
$s_j$ is the total number of tokens in $j^{th}$ section.
The tags of section and subsection from which a particular token originated are recorded.
These tags are passed through tag embedding module shown in lower right of Fig. \ref{scheme}.
This module computes the embedding representations of both tags and outputs the mean of the embedded tags.
The embedded tag representation is subtracted from the mean of the embedded token representations in each section.
This operation is performed by combining module indicated as $\mathcal{C}$ in Fig. \ref{scheme}.
The final embedding representation of the tokens in ${k_t}^{th}$ subsection of $j^{th}$ section, $F_{k_{t}}^j$, is obtained as follows,
\begin{equation}
    F_{k_{t}}^j =\left[ \frac{1}{s_j^{k_t}}\sum_{i=1}^{s_j^{k_t}}E_{i}^{j,k_t}\right] - G_{k_t}^{j}
\end{equation}
where $G_{k_t}^{j} = \frac{1}{2}\left[Em(Tag_j)+Em(Tag_{k_t})\right]$ is the embedded tag representation for $j^{th}$ section and its ${k_t}^{th}$ subsection, and $s_j^{k_t}$ is the total number of tokens in ${k_t}^{th}$ subsection of $j^{th}$ section.
Tag representation is subtracted from token representations due to possible inconsistencies in the information presented in FLPs and CLPs.
The selected terms for a particular section or subsection are pertinent to that section or subsection.
In this way, we discard a portion of the meaning of the words that are shared by all profiles and only focus on the remaining content that gives additional discriminative characteristics for differentiating LLPs from FLPs and CLPs.
The section and subsection tags used in SSTE are listed in Table \ref{tab:tags}. 
Fig. \ref{fig:con} shows the concatenated and cleaned textual information of the CLP shown in Fig. \ref{fig:profiles}.
The section and subsection tags for each word are shown by the same color as the word highlighted.
In order to classify LinkedIn profiles, the document embedding representation is computed as the mean of the final embedding representations.
The document representation is then passed into a binary classifier differentiating LLPs from FLPs and CLPs.

\begin{table}[t]
  \caption{List of section and subsection tags used in SSTE method.}
  \label{tab:tags}
  \resizebox{0.8\columnwidth}{!}{%
  \begin{tabular}{cc|cc}
    \toprule
    \textbf{Section tag} & \textbf{Subsection tags} & \textbf{Section tag} & \textbf{Subsection tags}\\
    \midrule
    Introduction&workplace, Location&Projects&Title, Date, Description\\
    Overview (About)&Description&Publications&Title, Journal, Description\\
    Experiences&Workplace, Role, Duration, Location, Description&Courses&Courses\\
    Educations&Institute, Degree, Duration, Description&Skills&Skills\\
    Licenses&Title, Company, Description&Scores&Test, Information\\
    Volunteers&Role, Organization, Duration, Description&Languages&Languages\\
    Honors&Award, Information, Description&Organizations&Organization, Role\\   
  \bottomrule
\end{tabular}
}
\end{table}
\section{Experiments and evaluations}
In this section, we conduct several experiments to evaluate our proposed SSTE method.
We utilized five different binary classifiers including LR, RF, SVM with linear, polynomial, and radial basis function kernels. The average value of accuracies and Fl-Scores obtained from all classifiers has been reported as the evaluation metrics in all experiments. 

\subsection{Evaluating the SSTE compared to the baseline for discrimiating LLPs from FLPs} \label{sec:numericalproposed}
The work done in \cite{adikari2020identifying} is chosen as the baseline for comparison with our proposed method.
To be fair, we exclude the number of connections, and the number of recommendations from the feature set proposed in \cite{adikari2020identifying} in order to compare the effectiveness of the methods to spotting the newly registered fake accounts.
We use $420$ and $420$ LLPs and FLPs for training.
The trained classifier is tested on $180$ and $180$ unseen LLPs and FLPs. 
Table \ref{tab:numericalproposed} shows the results for the baseline and the SSTE method.
In all embedding methods, SSTE outperforms the baseline.
BERT embedding has the best performance among all embeddings and improves the average accuracy by $17.79 \%$ compared to the baseline. 
In addition, the results for section tag embedding (excluding subsection tag embedding) is shown in Table \ref{tab:numericalproposed} as STE method.
In all embeddings, STE has lower performance compared to SSTE method.
One possible explanation is that by excluding subsection embeddings, the different subsections of one section is treated equally. 
In other words, although section embedding can introduce discriminative characteristics  between same subsection titles in different sections (e.g, "duration" in "experiences" and "educations"), excluding subsection embeddings will result in subtracting same embedding representation from different subsections of one section (e.g, "institute" and "duration" subsections in "educations") leading to introducing lower discriminative characteristics and hence, lower performance in STE compared to SSTE. 

\subsection{Comparison of textual data and numerical data}
We test the effectiveness of textual data compared to numerical data in identifying fake LinkedIn accounts. 
We only use embedding representations of concatenated textual data without using section and subsection tag embeddings.
The number of LLPs and FLPs used as training and testing set is the same as \ref{sec:numericalproposed}.
The results are shown in Table \ref{tab:numericaltextual}.
Comparing these results with the baseline results in Table \ref{tab:numericalproposed} shows that using textual data over numerical data results in a significant improvement (about $14\%$ on average for all embeddings).
In addition, we combine the numerical and textual data by concatenating them, and train our classifier using the new representations.
The results for combined data are shown in Table \ref{tab:numericaltextual} under "Numerical+Textual data" column.
The accuracy in three out of four embeddings is slightly improved.
The minor improvement can be accounted for by the fact that textual data can reflect the discriminative characteristics presented in numerical data by means of its length.
\begin{table}[t]
  \caption{The results of the baseline model \cite{adikari2020identifying}, section tag embedding (STE) method, and section and subsection tag embedding (SSTE) method in terms of average accuracy and F1-score for discriminating LLPs and FLPs. The training set contains $420$ LLPs and $420$ FLPs, and the testing set contains $180$ LLPs and $180$ FLPs.}
  \label{tab:numericalproposed}
  \resizebox{0.8\columnwidth}{!}{%
  \begin{tabular}{cccccccccc}
    \toprule
    Metric & \textbf{Baseline} &\multicolumn{4}{c}{\textbf{STE}} & \multicolumn{4}{c}{\textbf{SSTE}}\\
    
    & & GloVe & Flair & BERT & RoBERTa & GloVe & Flair & BERT & RoBERTa\\
    \midrule
    Avg. Accuracy (\%) &81.78 &87.78 &87.45 &94.28 &93.33 &94.78&94.17&96.33&95.00\\
    Avg. F1-score &0.816 &0.878 &0.875 &0.942 &0.934&0.947&0.941&0.963&0.950\\
  \bottomrule
\end{tabular}
}
\end{table}
\begin{table}[t]
  \caption{The results of using textual data and combined textual and numerical data in terms of average accuracy and F1-score for discriminating LLPs and FLPs. In both settings, section and subsection tag embedding are not utilized. The training set contains $420$ LLPs and $420$ FLPs, and the testing set contains $180$ LLPs and $180$ FLPs}
  \label{tab:numericaltextual}
  \resizebox{0.8\columnwidth}{!}{%
  \begin{tabular}{ccccccccc}
    \toprule
    Metric & \multicolumn{4}{c}{\textbf{Raw textual data (without SSTE)}} & \multicolumn{4}{c}{\textbf{Numerical+Textual data}}\\
   
     & GloVe & Flair & BERT & RoBERTa & GloVe & Flair & BERT & RoBERTa\\
    \midrule
    Avg. Accuracy (\%) & 92.78&90.89 & 95.55&93.61&93.56&90.39&95.67&93.83\\
    Avg. F1-score &0.926 &0.908 &0.955&0.937&0.936&0.903&0.956&0.939\\
  \bottomrule
\end{tabular}
}
\end{table}
\subsection{Evaluating the effectiveness of SSTE trained on LLPs and FLPs for detecting CLPs}
\begin{table}[t]
  \caption{The results of detecting CLPs using SSTE model trained on $600$ LLPs and $600$ FLPs in terms of average accuracy and F1-score. The trained SSTE is tested on $1200$ CLPs and $1200$ LLPs.}
  \label{tab:withoutclp}
  \resizebox{0.5\columnwidth}{!}{%
  \begin{tabular}{ccccc}
    \toprule
    Metric & GloVe & Flair & BERT & RoBERTa\\
    \midrule
    Avg. Accuracy (\%) & 71.43&75.88 & 72.26&76.12\\
    Avg. F1-score & 0.63&0.699 & 0.632&0.701\\
  \bottomrule
\end{tabular}
}
\end{table}
To assess the performance of our proposed method on detecting LLM-generated profiles as the next generation of fake accounts, we train the SSTE model on $600$ LLPs and $600$ FLPs, and then test it on unseen $1200$ CLPs and $1200$ LLPs.
Table \ref{tab:withoutclp} shows the results.
It can be seen for all embeddings that the accuracy is above $70\%$ showing that SSTE is able to spot some of CLPs as fake accounts despite the fact that it was not exposed to any samples of CLPs during training.
This is of important findings because in the near future many LLM rivals will be introduced, and this fact makes recognizing fake profiles generated with different LLMs extremely challenging.

\subsection{Using CLPs as fake accounts for training}
\begin{table}[t]
  \caption{The results of using CLPs as fake profiles in the training phase to identify FLPs in the testing phase in terms of average accuracy and F1-score. We use $1200$ CLPs as fake samples and $1200$ LLPs in the training phase, and evaluate the trained SSTE model on recognizing $600$ LLPs from $600$ FLPs.}
  \label{tab:clptrain}
  \resizebox{0.5\columnwidth}{!}{%
  \begin{tabular}{ccccc}
    \toprule
    Metric & GloVe & Flair & BERT & RoBERTa\\
    \midrule
    Avg. Accuracy (\%) &69.48 &57.27& 65.39 &60.38\\
    Avg. F1-score &0.564 &0.255 &0.472 &0.348\\
  \bottomrule
\end{tabular}
}
\end{table}
The most challenging task in collecting dataset for detecting fake LinkedIn accounts is to find fake profiles in this OSN. 
Therefore, in this experiment we use $1200$ CLPs as fake samples and $1200$ LLPs in the training phase, and evaluate the trained SSTE model on recognizing $600$ LLPs from $600$ FLPs.
The results are presented in Table \ref{tab:clptrain}.
The average accuracy among all embeddings is $63.13 \%$.
The obtained accuracy surpasses that of a random classifier, indicating that the model is effective in making predictions.
While there may be room for improvement, the results suggest that the model is performing reasonably well and is a promising starting point for further refinement.

\subsection{Determining the optimal number of CLPs for effective model training}\label{clpcount}
\begin{figure}[t]
  \centering
  \includegraphics[width=0.55\linewidth]{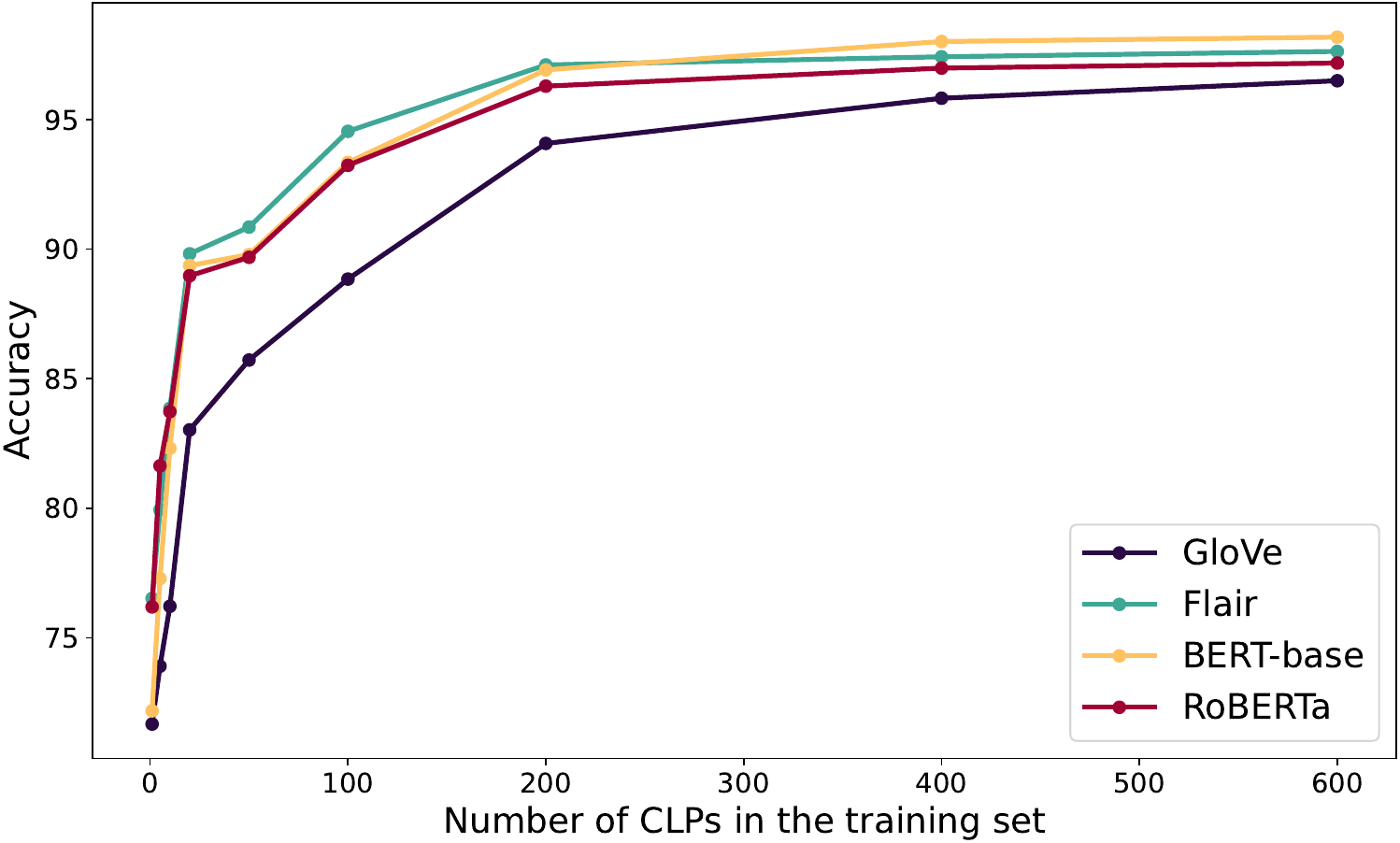}
  \caption{Performance of the SSTE on identifying CLPs from LLPs based on the number of CLP samples in the training set. The horizontal axis represents the number of CLPs used in the training set and the vertical axis represents the average accuracy on unseen data.}
  \label{Fig:nclp}
  \Description{}
\end{figure}
It is believed that the number of LLMs will expand rapidly in the near future.
In order to be able to identify profiles created with different LLMs in the OSN, we conduct an experiment to determine the optimal number of CLPs required in the training phase to obtain a model that can effectively detect CLPs in addition to FLPs.
We aim to investigate the impact of different embeddings on model performance, with a particular focus on the number of CLPs required for each embedding to achieve a satisfactory level of accuracy.
We use a training set including $n$ number of CLPs, where $n$ is incrementally increased from $1$ to $600$.
The number of LLPs and FLPs in the training set is set to $(600{+}n)$ and $600$, respectively.
The trained SSTE model is then tested on $(1200{-}n)$ and $(1200{-}n)$ LLPs and CLPs, respectively.
Fig. \ref{Fig:nclp} shows the results where horizontal axis represents the number of CLPs used in the training set and vertical axis represents the average accuracy obtained from testing set.
Our results indicate that the number of CLPs required for training varied depending on the type of embedding used.
Flair, BERT, and RoBERTa require only a small number of CLPs (around $20$) to achieve an accuracy of approximately $90\%$.
On the other hand, GloVe requires a larger number of CLPs to achieve a similar level of accuracy, but eventually converges to the performance of the other three embeddings.
It is possible that this difference is related to the nature of the embeddings themselves. Flair, BERT, and RoBERTa are contextual embeddings, which may be better suited for capturing the nuances of CLPs as they have been generated using GPT v3.5.
In contrast, GloVe is a static embedding, which may require more examples to learn the necessary features.
In the context of the increasing availability of LLMs, our study highlights the importance of optimizing model training with a minimal number of CLPs to effectively detect profiles created by different LLMs in addition to FLPs. 
Furthermore, our results suggest that the choice of embedding plays a vital role in the number of CLPs required, with contextual embeddings potentially requiring fewer samples for highly effective model performance.

\begin{figure}[h]
    \centering
    \includegraphics[width=0.65\textwidth]{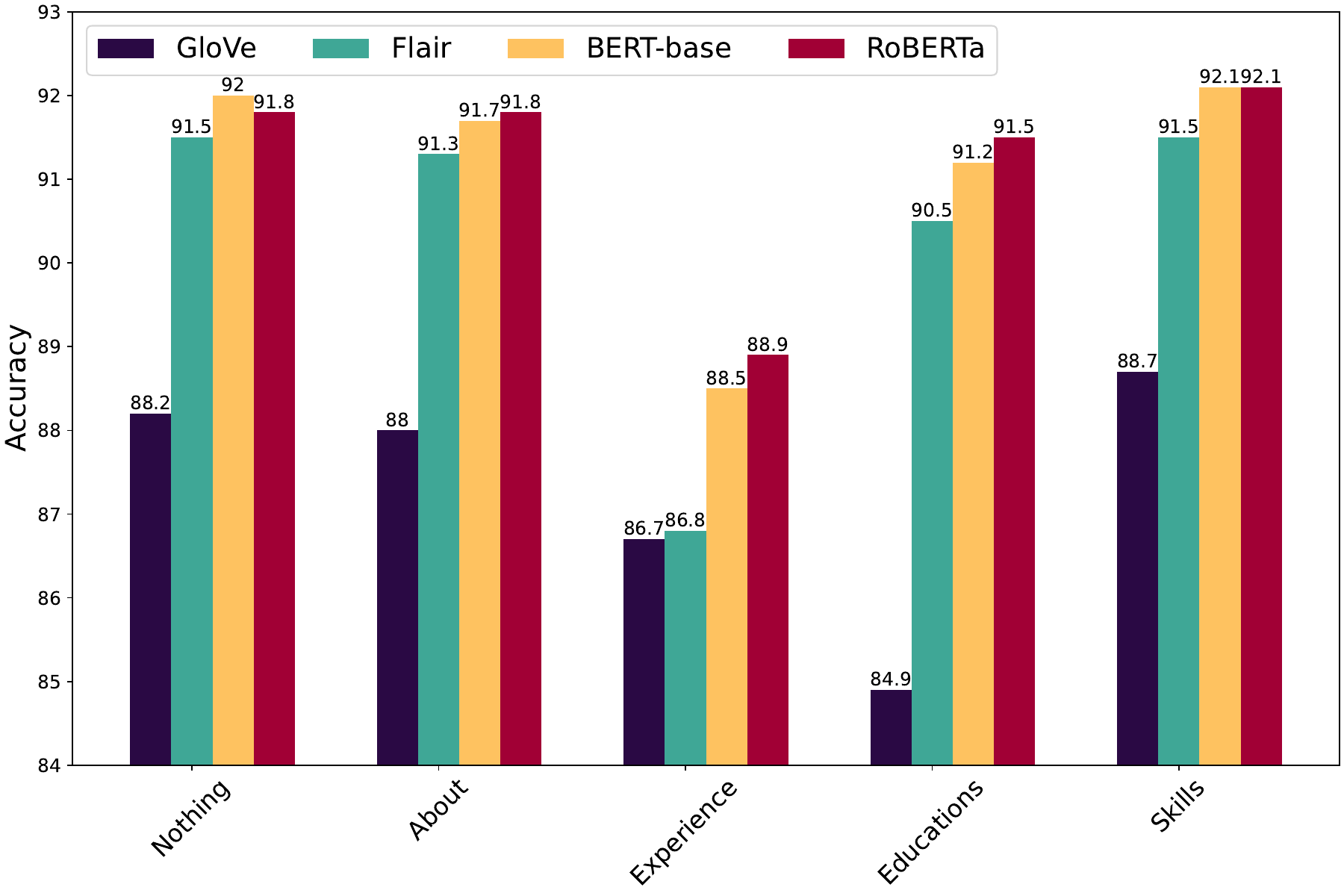}
    \caption{The impact of removing the textual data of about, experience, education and skill sections, which are the ones that LinkedIn members typically fill out the most, on the performance of SSTE. The baseline performance of SSTE where no section has been removed from the input is shown with "Nothing" label.}
    \label{fig:remove}
\end{figure}

\subsection{Evaluating the impact of LinkedIn profile sections on the model performance}

We conduct an assessment of the contribution made by different sections to the accuracy of the SSTE in differentiating LLPs from FLPs and CLPs.
$78.04\%$, $98.62\%$, $94.63\%$, $32.42\%$, $31.33\%$, $86.17\%$, $8.04\%$, $12.88\%$, $9.08\%$, $21.04\%$, $1\%$, $27.67\%$, and $19.71\%$ of LinkedIn users, in our dataset, filled out the About, Experiences, Educations, Licenses, Volunteers, Skills, Projects, Publications, Courses, Honors, Scores, Languages, and Organization sections, respectively.
To evaluate the effect of each section on the accuracy, we systematically remove the textual data of one section at a time while retaining the data of remaining sections.
The training set consists of $500$ LLPs, $480$ FLPs, and $20$ CLPs (as obtained in \ref{clpcount}), while the test set contains $240$, $120$, and $120$ unseen LLPs, FLPs, and CLPs, respectively.
The results, as illustrated in Fig. \ref{fig:remove}, indicate that leaving out the Experience section has the most impact on the model's performance compared to other sections.
However, in other cases (including sections that could not be presented in Fig. \ref{fig:remove} due to limited space), the performance changes are found to be negligible. 
This robustness demonstrates that the suggested method can effectively cope with situations where data from a section is unavailable.
Identifying the information provided by the Experience section that set it apart from other sections would require further research and analysis. 
Thus, we leave the investigation of these features and their impact for future research endeavors.

\section{Conclusion and future work}
We presented SSTE, a method for discriminating legitimate profiles from current fake and next-generation fake profiles (LLM-generated profiles) immidiately after registration in LinkedIn OSN.
Due to scarcity of data for LinkedIn fake account detection, we collected a dataset containing $3600$ profiles, including legitimate, fake, and ChatGPT-made LinkedIn profiles to conduct our experiments. 
In order to improve the discriminative characteristics of textual data provided in various sections of a LinkedIn profile, we merged section and subsection tags with these data by making use of a variety of word embeddings.
We compared SSTE with numerical-attribute based approaches and showed that it significantly outperformed these methods.
In addition, it was demonstrated that the SSTE is able, to some extent, to recognize LLM created profiles when the training set does not contain any LLM-generated profiles.
We further determined the minimal number of CLPs required in training set to achieve a significant accuracy on identifying LLM-generated profiles.
The results showed that SSTE required only $20$ LLM-generated profiles to be trained on in order to have an accuracy of about $90\%$.
This finding implies that with the emergence of abundant number of LLMs in the near future, SSTE can still detect fake profiles created by various LLMs accurately.

One significant aspect that is still unexplored in our present study is the use of LLMs to assist individuals in crafting sections of their LinkedIn accounts.
As a potential avenue for future research, it is vital to delve into the discerning features that can effectively distinguish between fake accounts that are entirely generated by LLMs and legitimate accounts that leverage LLMs to enhance particular portions of their profiles.

\section{Acknowledgments}
This research was supported in part by grant ARO W911NF-20-1-0254.
The views and conclusions contained in this document are those of the authors and not of the sponsors.
The authors also would like to acknowledge the important contribution made by David Chamberlin, Steve Elliott, and other LinkedIn users who helped us in the collection of the dataset used in this research.
\bibliographystyle{ACM-Reference-Format}
\bibliography{sample-base}


\end{document}